\pdfoutput=1

\documentclass[twocolumn]{paper}

\usepackage{natbib}
\usepackage{hyperref}
\usepackage{url}
      
\usepackage{graphicx}
\usepackage{subfig}

\usepackage[flushleft]{threeparttable}

\usepackage{asymptote}

\usepackage{amsmath}
\usepackage[mathscr]{euscript}
\usepackage{bm}

\usepackage{mathtools}
\DeclarePairedDelimiter{\ceil}{\lceil}{\rceil}

\usepackage[margin=1in]{geometry}
\usepackage[varg]{txfonts}

\usepackage{color}

\usepackage{fancyhdr}
\begin{document}

\title{Estimating the Handicap Effect in the Go Game:\\
       A Regression Discontinuity Design Approach}
\author{Kota Mori \thanks{
         {\tt kota.mori@aya.yale.edu}}
       }

\maketitle

\begin{abstract}
This paper provides an estimate for the handicap effect in the go game,
a board game widely played in Asia and other parts of the world.
The estimation utilizes a unique handicap assignment rule of the game,
where the amount of handicaps changes discontinuously 
with the players' strengths. 
A dataset suitable for this estimation strategy is collected
from game archives of an online platform.  
The result implies that an additional handicap typically 
changes the game odds by about 30 percent points, 
while the impact varies across the handicap level.
\end{abstract}


\section{Introduction}

Handicapping is a practice frequently applied in competitive games and sports
for the purpose of equalizing the players' chances of winning.
When the abilities of players differ so much that 
competition under the standard setup is considered ineffective,
certain disadvantage is imposed on the more skilled player, so that
the game odds become close to even.  
With appropriate amount of handicaps, diverse levels of contestants can participate
in the same tournament, or tutoring sessions become more effective.

An effective assignment of handicaps requires reliable prediction of their impact.
However, statistical inference on the handicapping impact is often challenging
since it by nature suffers from selection bias;
People play handicapped games {\it because} their ability gap is large.
Hence players playing even games are not necessarily comparable with 
those playing handicapped games.
Naive comparison of games with and without handicaps would produce 
a biased estimate since it cannot filter the
effect of handicaps from that resulted from other covariates.

This paper is an attempt to overcome this challenge in the case of 
the go game.
Go is a two-player board game that originates in ancient China.
It belongs to the class of two-player zero-sum games with complete information.
The game is played widely in East Asia, such as China, Korea Republic,
Japan, and Taiwan, while it has been gaining popularity in other areas of the 
world as well.

\begin{figure}[h]
  \begin{center}
    \begin{tabular}{ccc}
      \includegraphics[width=.4\hsize]{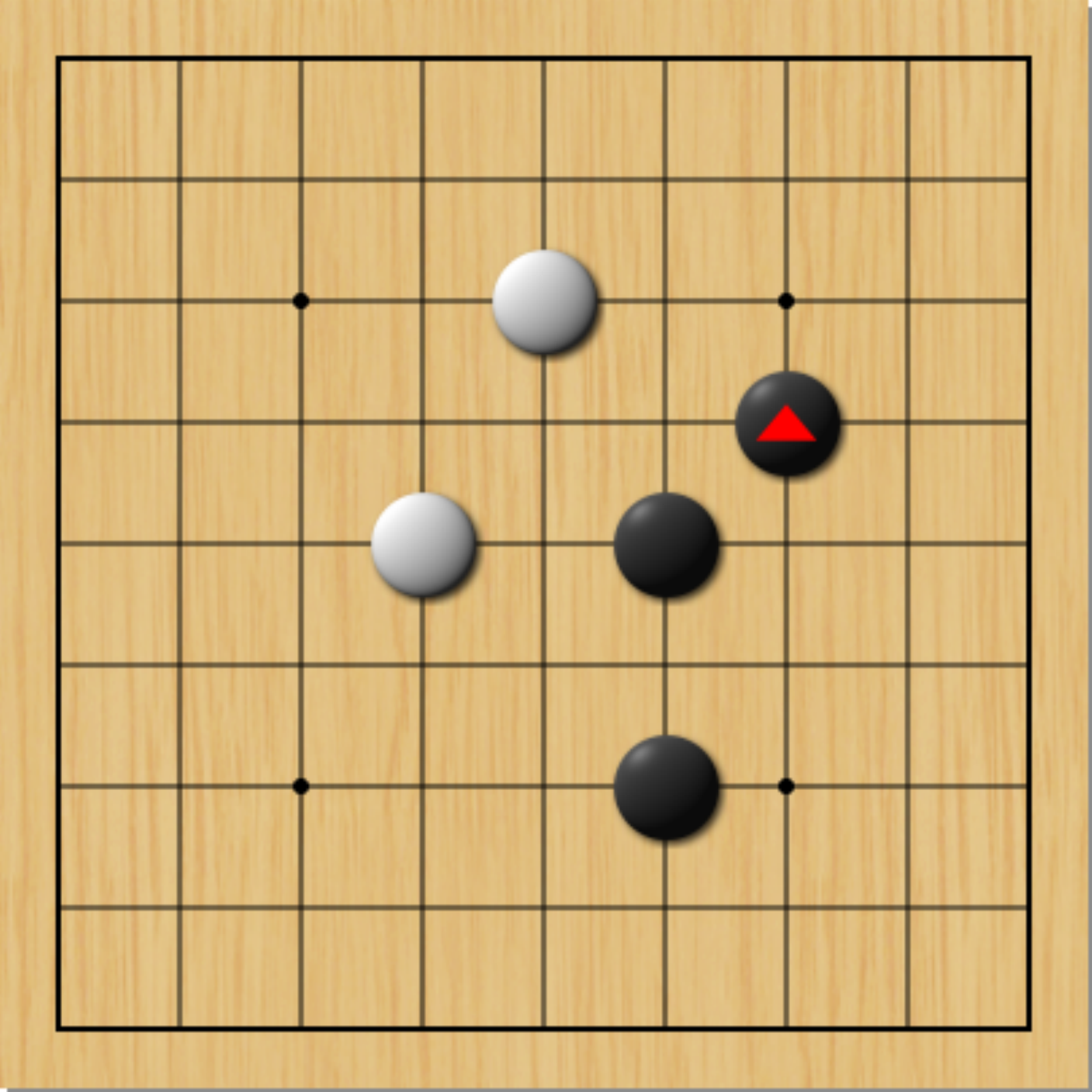} & 
      \;\;\; &
      \includegraphics[width=.4\hsize]{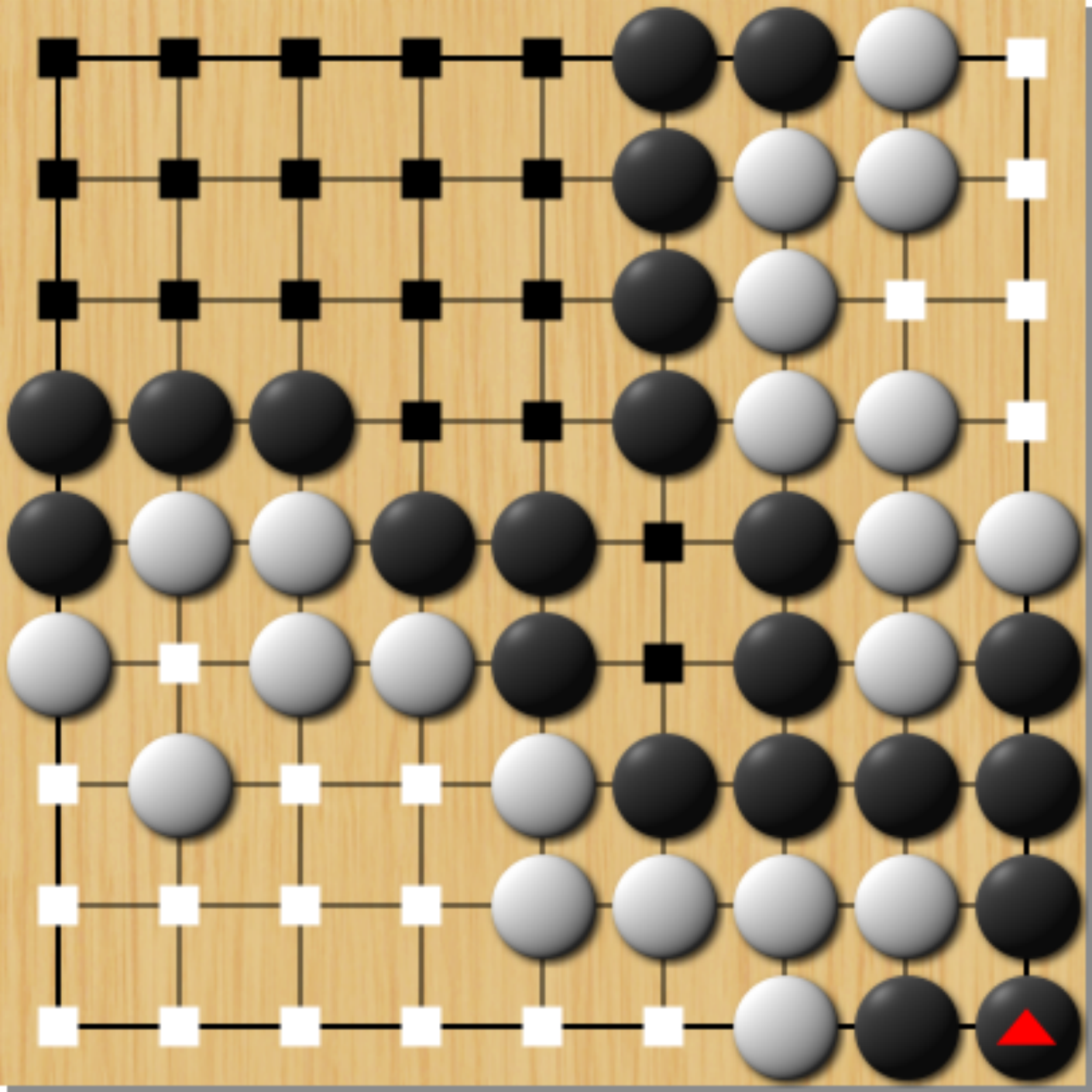} 
    \end{tabular}
    \caption{Left: 9 by 9 go board. Stones are placed at intersections. 
      Right: Example of end game and territory counting.  
      Both the black and white have 19 points.
      The images are generated using the software mugo (\protect\url{https://code.google.com/archive/p/mugo/})}
    \label{fig:goexample}
  \end{center}
\end{figure}

Go is played on a board with grids.
The official grid size is 19 by 19,
while smaller sizes are also used occasionally.
In a game, players hold either black or white stones and 
place them at intersections of the grids by turn (See the left panel of
Figure \ref{fig:goexample}). 
The goal of the players is to enclose a larger territory, 
which is defined as the number of empty intersections surrounded by stones of 
either color.
See the right panel of
Figure \ref{fig:goexample} for an example of end game configuration and territory counting.

In the official rule, the player who holds black makes the first move. 
Since he always has the same or a larger number of 
stones on the board than the white does,
it is known that he enjoys a certain advantage throughout the game.
To offset this first mover advantage, the white player is granted 
6.5 (or 7.5 in Chinese rule) points at the end of the game.
These additional points are called {\it komi}.

Handicapping system of the go takes advantage of the first mover advantage.
If the skill levels of two players differ significantly, 
then the more skilled one holds white (hence becomes the second mover) 
without receiving komi.
Hence, the black can fully enjoy the first mover advantage, which would 
close the skill gap to some extent.
This setup is called a {\it sen} game.

\begin{figure}[h]
  \begin{center}
    \begin{tabular}{ccc}
      \includegraphics[width=.4\hsize]{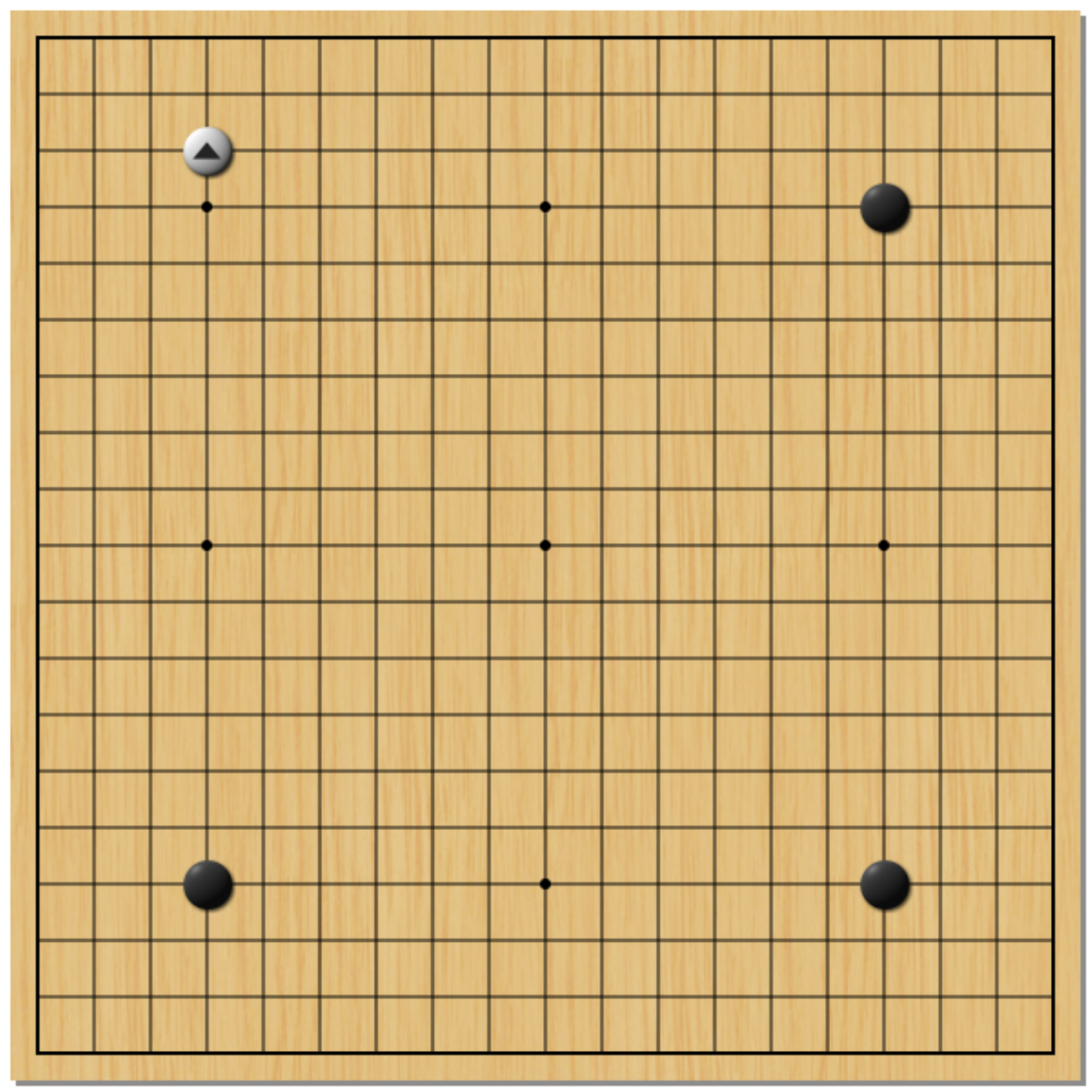} & 
      \;\;\; &
      \includegraphics[width=.4\hsize]{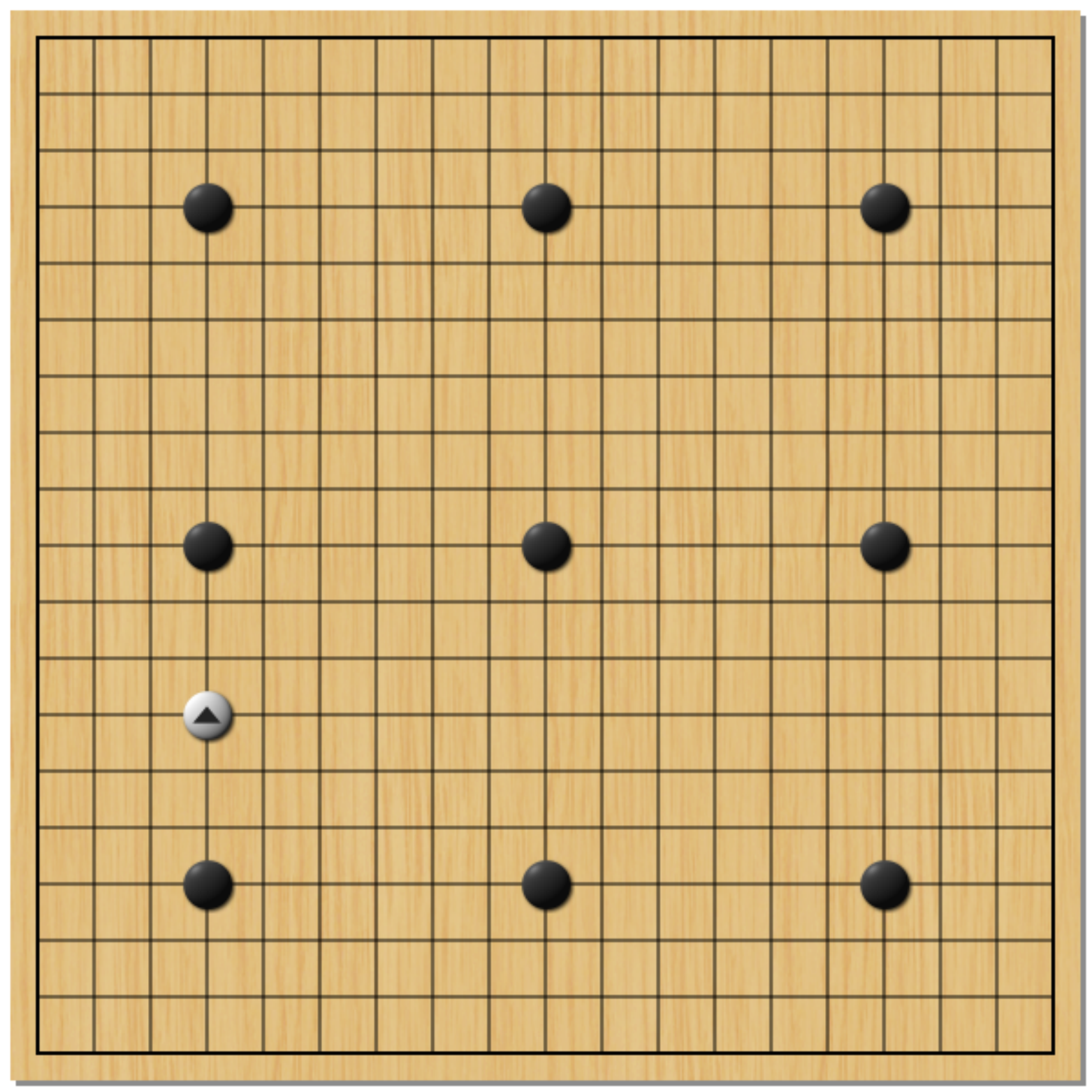} 
    \end{tabular}
    \caption{Left: Opening of a three stone game.
             Right: Opening of a nince stone game.
     The images are generated using the software mugo}
    \label{fig:gohandi}
  \end{center}
\end{figure}

If the skill gap is so large that the first mover advantage is not 
sufficient to fill it, the less skilled is allowed to add another 
stone at the beginning, typically at designated positions called 
{\it stars}. 
This setup is called a two-stone game.
As the abilities are apart even further, the number of handicap stones 
increase to three, four and so on.
See Figure \ref{fig:gohandi} for opening configurations of handicapped games.

Handicap assignment is often determined by a discrete rating 
called {\it dan-kyu} system.
An absolute beginner is given a kyu with some large number such as 30 kyu
or 30k in short.
As his skill reaches certain thresholds, he gets promoted to 29k, 28k, and so on.
1k is followed by 1 dan or 1d, and 
the number starts to climb as 2d, 3d and so on. 
Typically 9d is the highest rating.

In many platforms, players with the same dan-kyu rating play even games, 
where their colors are determined randomly and the white player receives 
komi.
When the ratings of the players are different, 
the lower ranked player is entitled 
as many handicaps as the rating difference.
For example, 2k and 3k play sen games, where the 3k holds black 
with no komi given to the white.
9d and 1k, with rating difference of nine, play nine-stone games.

This assignment rule is discontinuous in players' skill level,
and hence forms a regression discontinuity design, providing a
quasi-experimental situation for causal inference on the handicap effect.
Suppose a 1d player plays two games, one against a ``weakest'' 1d, 
and the other against a ``strongest`` 1k.  
While the two opponents are similar in terms of the skill levels,
the setup would become even and sen respectively, due to their ratings.
Thus, the difference between the two games is arguably attributable
to the handicapping.

In this paper I apply this strategy to quantify 
the handicap effect on the game outcomes.
A dataset suitable for this approach is collected from game archives of
the KGS, an online go platform.

The rest of the paper is organized as follows.
Section \ref{sec:data} describes the data.
Section \ref{sec:method} formally states the estimation strategy.
Section \ref{sec:result} presents the results.
Section \ref{sec:concl} concludes.

\section{Data and Descriptives} 
\label{sec:data}

\subsection{Sampling players}

Data for this study have been collected from the archives of the KGS,
an online platform for the go game.\footnote{\url{http://www.gokgs.com/}}
Games played on the KGS are all stored on the server.
Unless the players want otherwise, records of all rated games are 
open to the public.\footnote{Rated games on the KGS are the games that influence
the players' future ratings. Other types include free games, teaching games, and
demonstration. This paper focuses on rated games.}
A game record is saved in a text file of the smart go format (SGF), 
which contains the actual move sequences and the outcome, as well as 
meta information such as the player IDs and ratings,
and the date and time when the game started.

As far as the author recognizes, the KGS does not provide
an API or quarry feature that allows for randomized sampling of the games;
The games are only viewable when searched by a player ID.
Hence, I first collected a list of active player IDs by 
manually recording the player IDs logging onto the 
``English game room.\footnote{A room is a community within the KGS.  
A room may be formed for a certain geographic area or language,
or used for a specific event.  ``English game room'' is one of the 
largest rooms.  
Despite the name, many non-native English speakers,
including Asians and Europeans, play in the room.}'' 
For the sake of geographical heterogeneity, the sampling has been conducted
at four different timings.\footnote{Samplings were conducted at 
5am, 11am, 5pm, and 11pm of May 3rd, 2014, 
in the Eastern Standard Time (UTC -5).}

I also added the 100 top rated players as of May 7th, 
2014 to the list.\footnote{\url{http://www.gokgs.com/top100.jsp}}
Top players on the KGS include professional players and 
amateurs close to that level. 
By including these players the findings of the analysis 
are more likely to be generalizable to the professionals.

\begin{figure}[h]
  \begin{center}
    \includegraphics[width=.9\hsize]{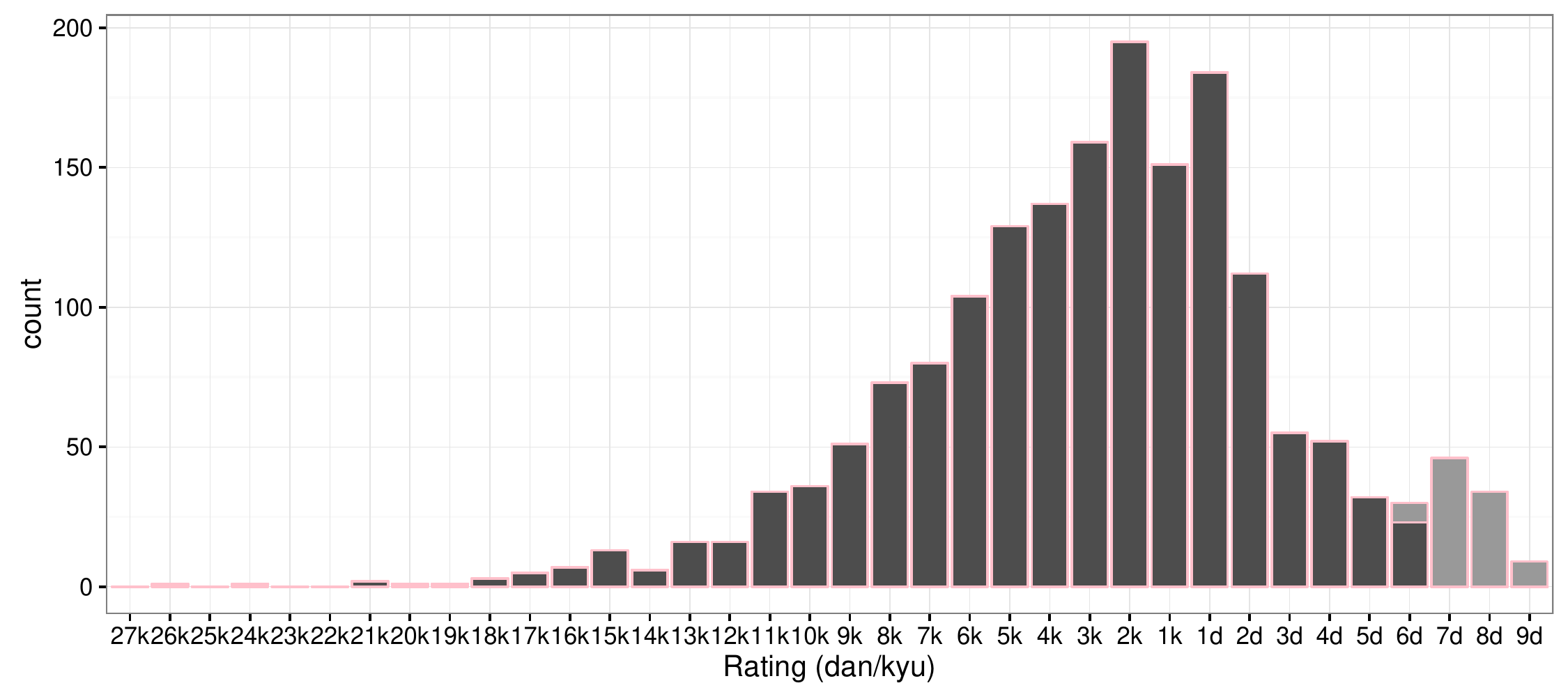}
    \caption{Distribution of the ratings in dan-kyu system.
      The darker bars indicates the number of sampled players, while 
      the lighter indicates the top rated players.
      This chart covers 1775 players who played at least one 
      rated game in the period between March 30, 2013 and May 13th, 2014.}
    \label{fig:ratedist}
  \end{center}
\end{figure}

The final list consists of 1940 players.
Among them, 1775 players (1679 sampled and 96 top-rated) are active 
in a sense that they played one or more rated games in the study period.
Figure \ref{fig:ratedist} shows the distribution of the player ratings in 
dan-kyu system.

\subsection{Data processing}

For each player, all rated games in the past four-hundred and ten days were collected
(from March 30, 2013 through May 13th, 2014).
Games before that was not collected because,
as described below, players' continuous rating is only 
available for the past four-hundred days.
Hence, collecting data of older games does not increase the size of the final dataset.
Each game record contains player IDs and their rating in the dan-kyu system, 
the date and time when the game started, the number of handicap
stones and komi, and the game outcome.

\begin{figure}[h]
  \begin{center}
    \includegraphics[width=.8\hsize]{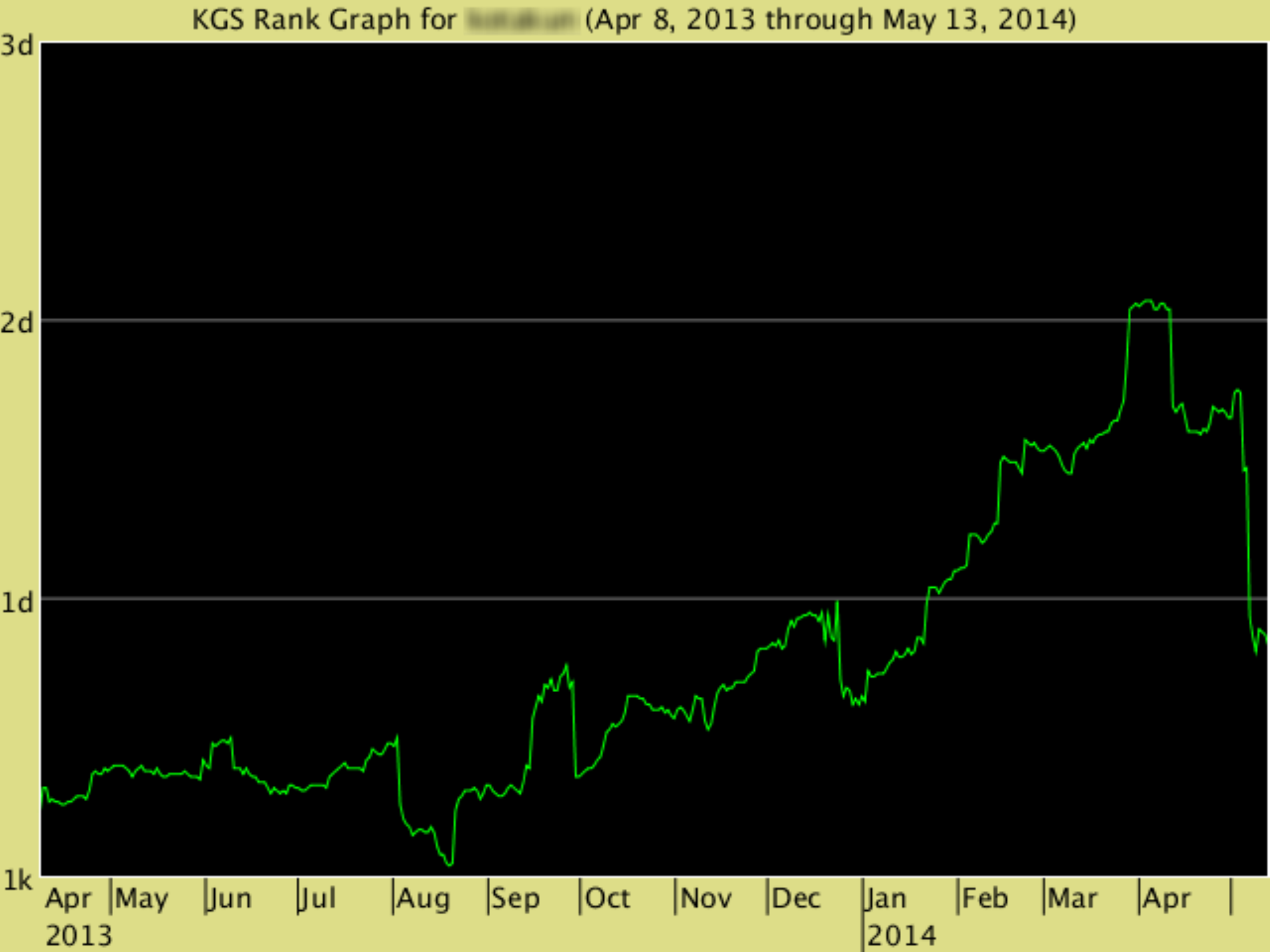}
    \caption{Example of a KGS rating plot.
      The rating history of the past 400 days is provided
      as a PNG image file.  Charts are updated daily.
      Player ID is made blur by the author.}
    \label{fig:kgsrank}
  \end{center}
\end{figure}

The process above yields a list of player IDs, including 
the sampled players as well as their opponents.
For each player, the image file of the rating history is downloaded
from the KGS archives, which shows a time series plot of the player's rating. 
The file indicates not only the discrete dan/kyu rating of the player, 
but also how high or low the player is rated within the same dan/kyu rating.
An example of the rating graph is shown in Figure \ref{fig:kgsrank}.
The graph is updated in daily basis, and covers at longest the 
period since 400 days ago through the present.

The pixel information of the image files were processed to recover the
time series of the players' ratings in a continuous scale.
The results are then merged with the game record data using the 
starting time of games as the key.

\subsection{Data filtering}

I removed the games where the continuous rating is missing
for a player or both from the analysis.
Missing values can occur if the ID has been deleted or inactive for 
a long period, and hence the rating graph is not available.
I also removed the games where the rating of a player or both is 
unstable, indicated by a question mark added to the discrete rating
({\it e.g.} 3k?).
This means that the player has not played as many games recently
as the KGS system can evaluate his or her rating 
at a certain credible level.

Further, I removed the games with a non-standard handicap setup. 
At default, the lower ranked player receives as many handicap stones 
as the difference in the dan/kyu ratings.
For example, players play even games  
(no handicap stone, the white receives 6.5 or 7.5 points as komi)
if they have the same discrete rating.
For one rank difference ({\it e.g.} 3d vs 4d, 1k vs 1d), games are 
played at the sen setting; 
The higher ranked player takes the white with only 0.5 points of 
komi.\footnote{For sen games on KGS,
the white is usually given 0.5 point of komi so as to reduce the chance of draw.}
For larger rank differences, the lower ranked player is entitled one handicap
stone per each rank deviation.  For example, 1d and 3d, 
two rank difference, play a two-stone game, where 3d takes the white.
Most of the games in the dataset follow these standard setup of the KGS. 
The games deviating from it are possibly arranged specifically between the
players, and thus removed from the analysis.

After applying the above filtering, 85 percent of the game records remained. 
The final dataset has 895,050 games with 47,348 unique players.

\subsection{Descriptives}

\begin{figure}[h]
  \begin{center}
    \includegraphics[width=.95\hsize]{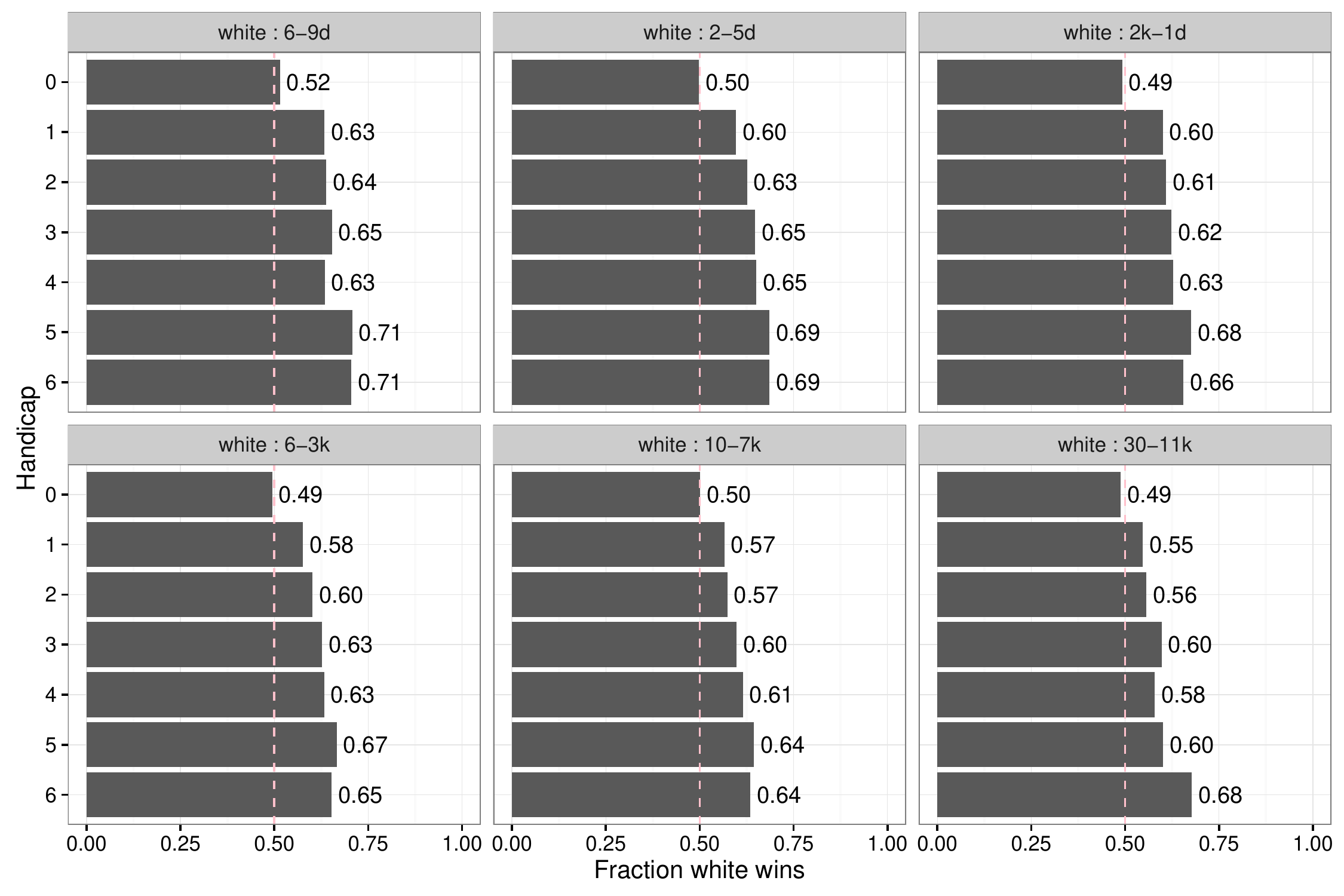}
    \caption{Fraction of games the white wins by the handicap level.
       The vertical axis indicates the handicap level;
       0 means even, 1 means sen games, 2 or larger numbers 
       indicate the number of handicap stones.
       Despite the disadvantageous setups, the white
       tends to win more in heavily handicapped games.}
    \label{fig:whitewin}
  \end{center}
\end{figure}

Figure \ref{fig:whitewin} shows the fraction of games the white wins 
by the handicap level.
The games are grouped by the discrete ratings of the white. 
Consistently across the player's levels, white players tend to win 
{\em more} in games with larger handicaps.
That is, the white wins more in setups where he or she plays at a more 
disadvantageous setting.

This unintuitive pattern can be understood as a result of 
selection bias. 
Those who play even games are not comparable with those who play
handicapped games since the game setup is chosen in accordance with their
skill differences.
Hence, direct comparison of the winning probability across the game setup 
does not necessarily reflect the impact resulted from the handicapping, but
may reflect the skill differences among those who play different setups.
In the present case, while handicaps may give advantages to the black, the 
impact may not be as large as it can equalize the game odds.
Causal inference on the effect of handicapping requires effective control over the
selection problem.

\section{Regression Discontinuity Design}
\label{sec:method}

The regression discontinuity design (RDD) is a framework to estimate 
causal impact of a binary treatment variable $X$ on some outcome 
measurement $Y$ \citep{Imbens2008}.
The RDD requires that there exists 
another variable $Z$, such that $X=1$ if and
only if $Z \ge c$, where $c$ is a known constant.\footnote{This is the
condition for so called the sharp RDD. The fuzzy RDD requires that the
probability that $X=1$ changes discontinuously at $Z=c$.
See \citet{Imbens2008} and \citet{Lee2010} for general discussions.}
$Z$ is called a running or forcing variable for $X$.
In case the treatment $X$ is not randomly assigned, the RDD provides a 
quasi-randomized experiment for the treatment effect. 
Provided that other covariates distribute continuously at $Z=c$,
the cases where $Z$ is marginally above the threshold (thus $X=1$) and 
the cases where $Z$ is marginally below (thus $X=0$) are quite similar
except that only the former receives the treatment.
Therefore this comparison allows one to estimate the causal impact of 
the treatment $X$.

In the context of this study, handicapping corresponds to the treatment, 
and its causal impact on the game outcome is of interest.
The running variable can be constructed by combining the ratings of
two players.
Let $d_i$ and $ r_i$ denote the discrete and continuous rating of 
the player $i$, where $i = W$ (white) or $B$ (black).
Without loss of generality, assume that $d_i$ is given by an integer 
defined by $d_i = \ceil{r_i}$, where $\ceil{x}$ indicates
the smallest integer greater than or equal to $x$. 
Then, for any nonnegative integer $H$, 
\[
d_W = d_B + H  \Longleftrightarrow 
H \le d_W - r_B < H+1
\]
This implies that $d_W-r_B$ is a running variable that uniquely 
determines the handicap level.
For example, a pair of players play an even game if and only if 
$d_W = d_B$, or equivalently, $0 \le d_W - r_B < 1$.
Similarly, a necessary and sufficient condition for the sen setup is 
$1 \le d_W - r_B < 2$, and 
that for two-stone game is $2 \le d_W - r_B < 3$.
Generally, let $H\in \{0, 1, 2, \dots\}$ denote the handicap level, 
where $H=0$ means even, $H=1$ means sen, and $H\ge 2$ indicates the
number of handicap stones.  Then, a necessary and sufficient condition 
for the setup $H$ is $H \le d_W - r_B < H+1$.  

In the estimation, the white's winning probability is set as the outcome measurement.
The causal effect of handicapping on the winning probability is estimated 
taking advantage of the RDD described above.
That is, the difference in the winning probability below and above 
$d_W - r_B = H$ in interpreted as the impact of switching the handicap level 
from $H-1$ to $H$.

The dataset includes cases where the running variable and the 
handicap level are inconsistent, that is, $d_W - r_B \not\in [H, H+1)$.\footnote{This
occurs for 25,926 games or 2.9 percent of the final dataset}.
This is because players' continuous ratings are measured with error, as they 
are recovered from pixel information of image files. 
Recent theoretical studies find that the mismeasured running would nullify the
standard polynomial regression estimates \citep{Davezies2014, Yanagi2014},  
and there has been no universal solution to this problem.
In this paper the treatment effect is estimated by only using the observations where
the running variable and the handicap level are consistent with each other. 
\citet{Yu2012} studies conditions under which this approach is valid.

\section{Results}
\label{sec:result}

\begin{figure}[h]
  \begin{center}
    \includegraphics[width=.95\hsize]{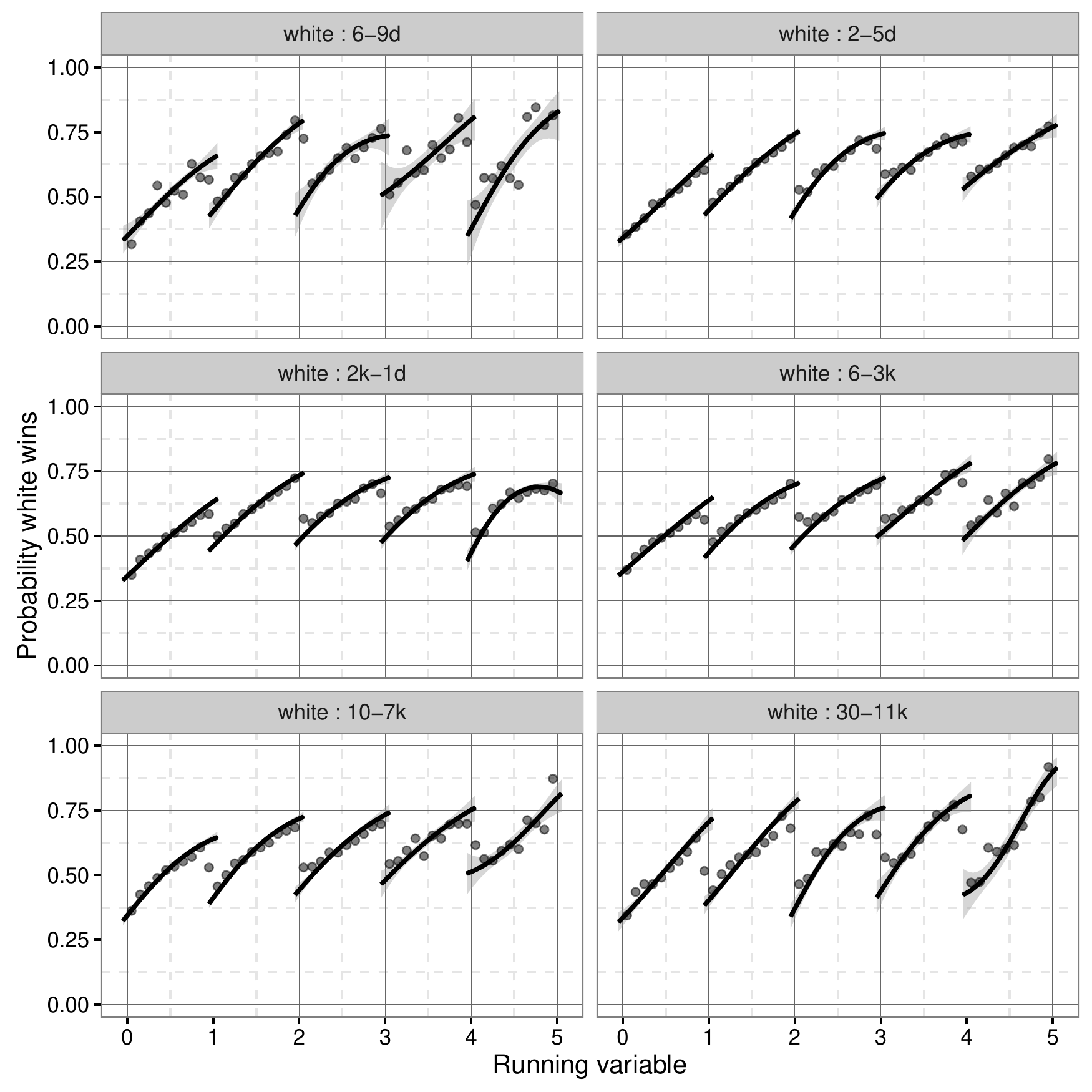}
    \caption{Predicted probability for the white to win.
      Each curve is estimated by 
      polynomial logistic regression of degree two.
      The gray bands represent the 95\% confidence intervals.
      The dots indicate the local averages.
      The drop in the probability at cutoff points can be 
      interpreted as the correponding handicap effect.}
    \label{fig:rdd}
  \end{center}
\end{figure}

Figure \ref{fig:rdd} presents the estimated winning probability of the white 
as a function of the running variable $d_W-r_B$.
The predictive line is estimated by  
polynomial logistic regression of degree two, separately for each
handicap level.  
The gray band shows the 95\% confidence interval.
Within the same handicap level, white's odds are increasing in the 
running variable. 
This is because the running variable, defined by $d_W-r_B$, 
indicates the relative strength of the white 
{\it vis-a-vis} the black.

At cutoff points, the white's winning probability drops substantially.
This reflects the impact of the handicap change, that is, 
the disadvantage that the white incurs by playing with 
an additional handicap as the opposing black player falls one rank
lower.

For even games, the predicted probability is about a half at the 
mid point of the range.
As the handicap level becomes large, however, the winning probability
tends to shift upward.
This is consistent with the observation that the white tends to win more
in handicapped games.
This implies that the handicapping tends to be insufficient 
for games where two players' skills are far apart, and
the game odds remain skewed towards the white.

\begin{figure}[h]
  \begin{center}
    \includegraphics[width=.95\hsize]{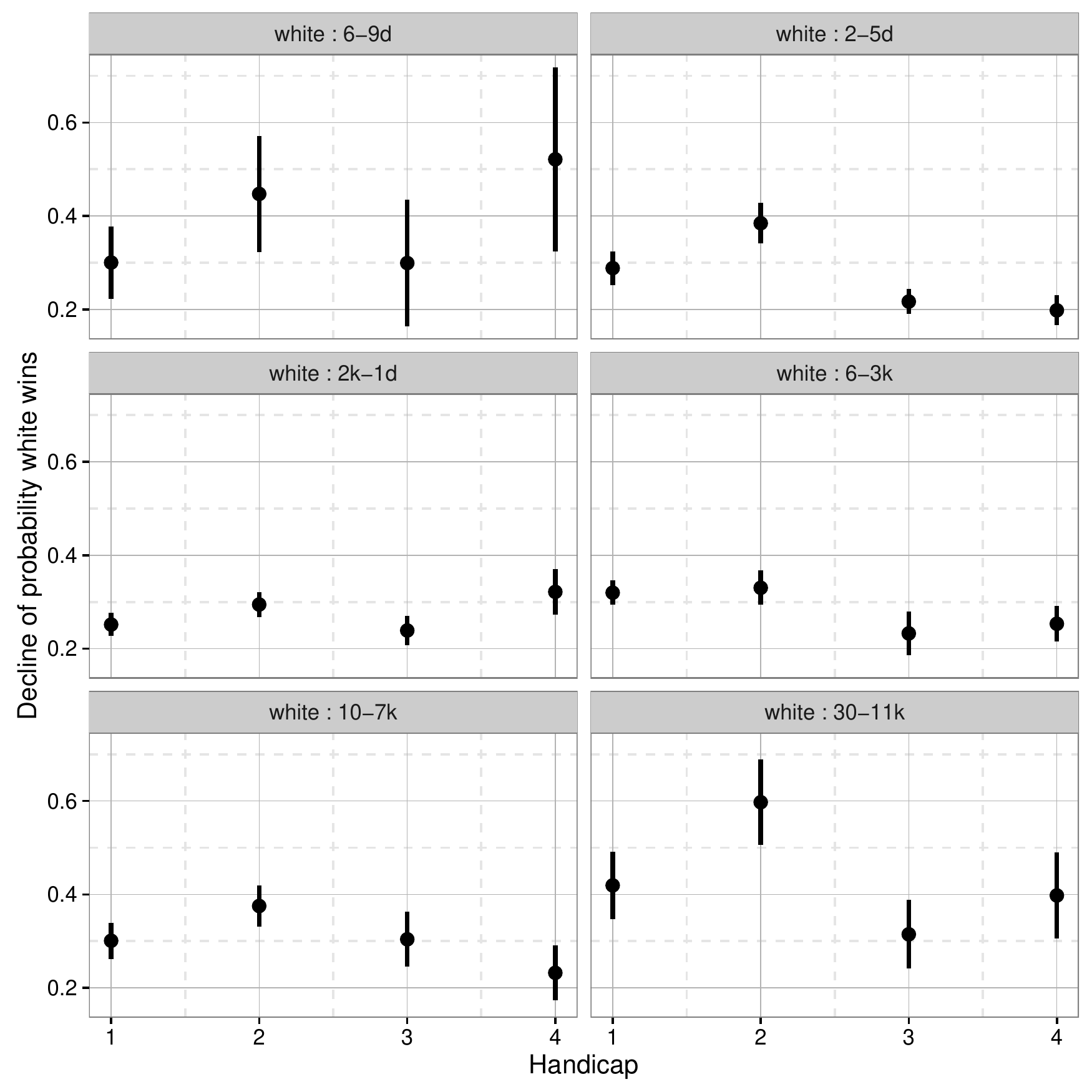}
    \caption{The handicap effect estimated by the local linear
      regression.  Bandwidths are selected by the criteria 
      suggested by \citet{Imbens2012}.  The dots and segments represent the
      point estimate and the corresponding confidence interval
      of the significance level 95\%.}
    \label{fig:estimate}
  \end{center}
\end{figure}

Figure \ref{fig:estimate} summarizes the estimated magnitude of handicap impact from the 
local linear regression at both below and above the 
cutoffs \citep{Imbens2008, Lee2010}.
The bandwidths are selected by the criteria suggested by \citet{Imbens2012}.
The estimation is conducted using the rdd package by \citet{Dimmery2013}. 
The dots and segments represent the
point estimate and the corresponding confidence interval
of the significance level 95\%.

The estimated impact of shifting from an even game to a sen game is
about 25 to 30 percent point.
This means, if the skill difference between a pair of players is such
that one of them wins 70 percent of the games or more in the even
setup, there is a good chance that removing komi realizes closer 
competition.

The impact is larger for the shift from sen to two-stone game, ranging
about 30 to 45 percent point.
Presumably advantage on the board due to the additional stone is more 
effective than the point advantage that realizes only at the 
end of the games.

Adding the third handicap stone brings smaller impact of about 20 to 30 percent point.
This can be because in both two-stone and three-stone games 
the white can assume one corner.
As the corners are considered valuable at the beginning, the two 
setups may not differ as much as sen and two-stone setups do.

The estimate for the impact of the fourth handicap stone is relatively
unstable due to the smaller sample size, ranging 20 to 50 percent point.
The impact seems about the same or slightly larger than that of the third
stone.

\section{Concluding Remarks}
\label{sec:concl}

This paper provides an estimate for the causal impact of handicaps in the go game,
taking advantage of the regression discontinuity design due to the unique
assignment rule of handicaps.
Typically handicaps reduces the probability that the white wins 
by about 30 percent point,
while variations are observed across the amount of handicaps.

Handicap assignments are often managed heuristically 
based on the experiences that have grown over the history.
Statistical evidence as presented in this paper will help complement the practice
and enhance more effective operation of games and competition. 

\bibliography{bib/kgsrd.bib}

\end{document}